\documentclass{iopart}

\usepackage{graphics,graphicx}

\usepackage{amsfonts,wasysym}

\usepackage{psfrag}
\usepackage{dcolumn}
\usepackage{psfrag}

\newcommand{\beq}{\begin{equation}}
\newcommand{\eeq}{\end{equation}}
\newcommand{\bea}{\begin{eqnarray}}
\newcommand{\eea}{\end{eqnarray}}

\newcommand{\hc}{{\sf h}}

\newcommand{\dtheta}{{\rm d}\theta}
\newcommand{\dphi}{{\rm d}\phi}

\begin{document}

\title{Status of black-hole-binary simulations for gravitational-wave
  detection} 

\author{Mark Hannam} 
\address{Physics Department, University College Cork, Cork, Ireland} 

\date{\today}

\begin{abstract}
It is now possible to theoretically calculate the gravitational-wave signal 
from the inspiral, merger and ringdown of a black-hole-binary system. The
late inspiral, merger and ringdown can be calculated in full general relativity
using numerical methods. The numerical waveforms can then be either 
stitched to inspiral waveforms predicted by approximation techniques (in
particular post-Newtonian calculations) that start at an arbitrarily low frequency,
or used to calibrate free parameters in analytic models of the full waveforms. 
In this review
I summarize the status of numerical-relativity (NR) waveforms that include
at least ten cycles of the dominant mode of the GW signal before 
merger, which should be long enough to produce accurate, complete 
waveforms for GW observations.   
\end{abstract}


\section{Introduction and overview}
\label{sec:intro}

A worldwide network of detectors~\cite{Abramovici92,Acernese2006,GEOStatus:2006}
is poised to make the first 
direct observation of gravitational waves (GWs). Gravity is the weakest of
the fundamental forces, and even GWs produced by the collision
of two black holes --- one of the most likely sources for the first 
detection --- will produce a signal obscured by detector noise;
signal-to-noise ratios (SNRs) less than ten may be typical.
One way to locate the signal is to compare the detector data against
a collection (template bank) of the signals predicted by Einstein's 
general theory of relativity for a range of potential sources. In the
case of black-hole-binary mergers, the only way to calculate the signals
predicted by the full theory is to solve Einstein's equations on a computer. 

Numerical solutions of Einstein's equations for the last orbits and 
merger of a black-hole binary, the ringdown of the single black hole 
that remains, and the GWs emitted in the process, became possible
in 2005~\cite{Pretorius:2005gq,Campanelli:2005dd,Baker05a}.
Since that time many simulations have been performed, producing
results relevant to mathematical general relativity, black-hole physics,
and galactic astrophysics. In this article, however, I review
the status of black-hole-binary simulations for the purpose of GW detection.

\begin{figure}
\centering
\includegraphics[width=60mm]{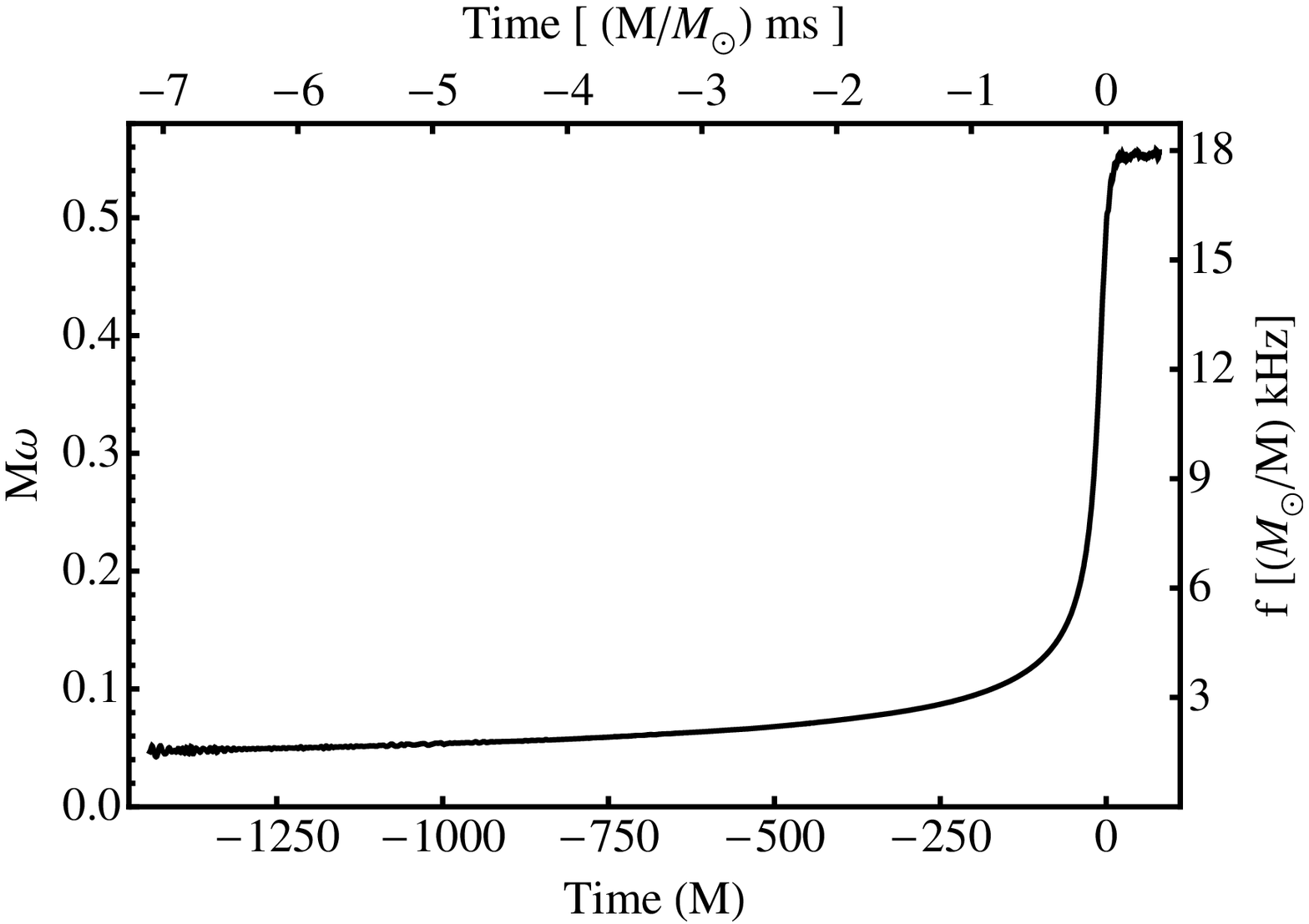}
\includegraphics[width=57mm]{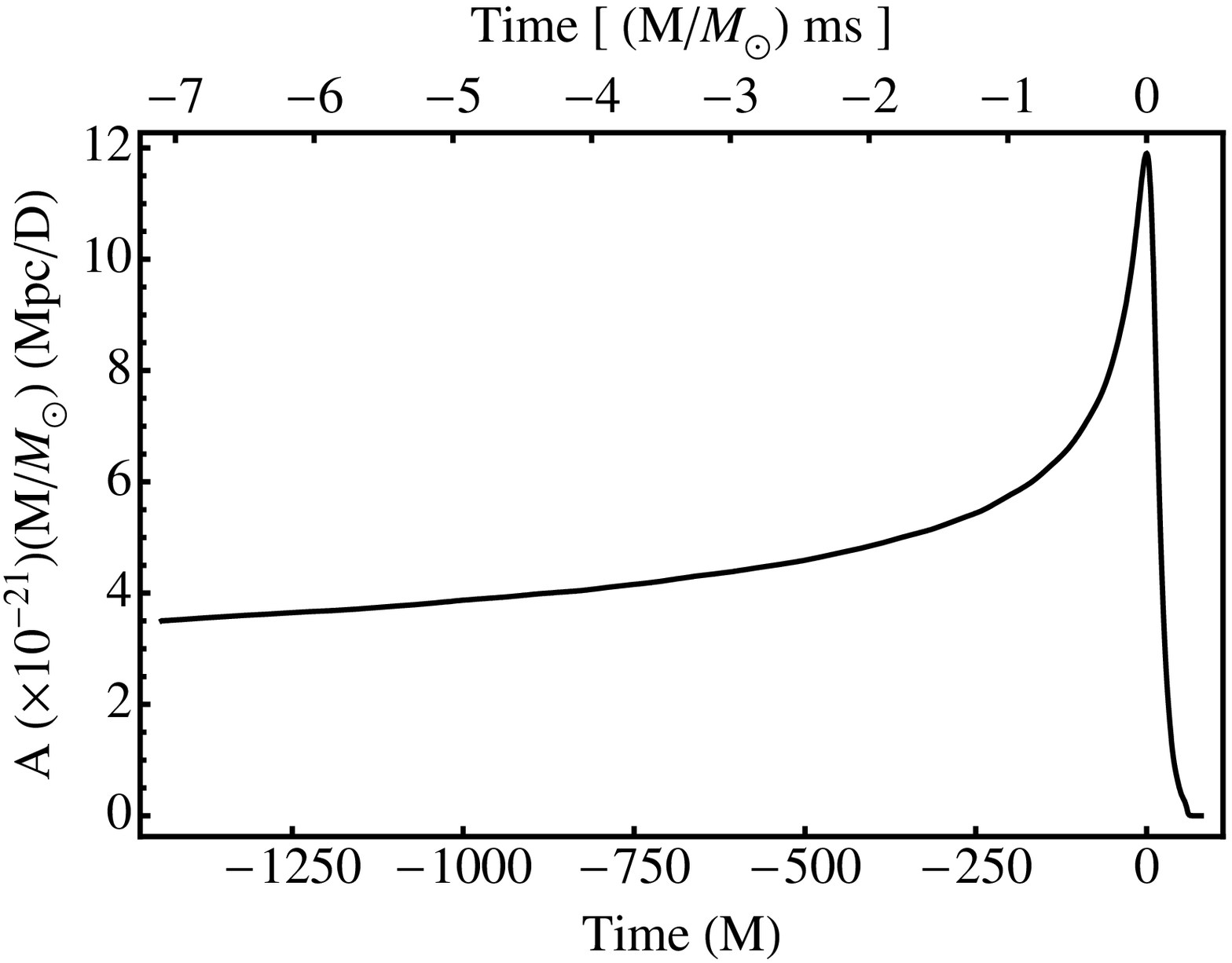}
\caption{Frequency and amplitude of the GW signal from a numerical simulation 
of the last eight orbits, merger and ringdown of an equal-mass nonspinning binary. 
The amplitude is given for an optimally-oriented binary. 
The time is given in units of the total mass $M$ of the
binary, and in milliseconds. The frequency is in dimensionless units $M\omega$
and kHz. The amplitude of the GW strain $\Delta L/L$ is dimensionless and scales
with respect to the binary's mass $M$ and distance $D$ (in megaparsecs) 
from the detector. (Data from high-resolution D12 simulation used
in~\cite{Hannam:2007ik}.)} 
\label{fig:FreqAmp}
\end{figure}

\begin{figure}
\centering
\includegraphics[width=70mm]{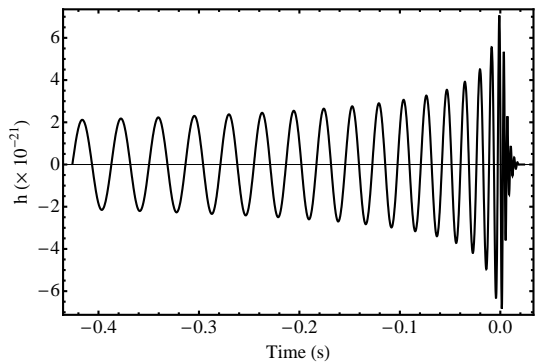}
\caption{The full GW strain signal for the same configuration as in Fig.~\ref{fig:FreqAmp},
for a 60\,$M_\odot$ binary located 100\,Mpc from the detector.}
\label{fig:waveform}
\end{figure}

Consider the archetypal black-hole-binary configuration studied in a numerical
simulation: two black holes of equal mass, with no spin, in orbit with zero
eccentricity. If the two black holes orbit with a frequency $\Omega(t)$, then the 
frequency of the dominant GW mode is $\omega(t) \approx 2 \Omega(t)$. 
As GWs are emitted,
the black holes spiral slowly inwards, and the GW frequency increases; so too 
does the GW amplitude, $A(t)$. The rate of inspiral grows, and the GW 
frequency and amplitude sweep up until the two black holes plunge together 
and merge, at which time the GW amplitude peaks and then decays exponentially, 
while the frequency levels off at the ringdown frequency of the remnant Kerr black 
hole. This behaviour is illustrated in Fig.~\ref{fig:FreqAmp} for the last sixteen 
cycles (eight orbits) and merger and ringdown from a numerical simulation 
of an equal-mass nonspinning 
binary. The time is shown both in units of the total mass $M$ of the binary 
(which is an overall scale factor for the numerical solution), and the
corresponding time in milliseconds for binaries with the total mass as some 
multiple of the mass of the sun, $M_\odot = 1.477 \times 10^{3}$\,m. As an
example of a complete waveform, Fig.~\ref{fig:waveform} shows the wave strain 
$h(t) = \Delta L(t)/L$ from a 60\,$M_\odot$ binary located 100\,Mpc away and 
optimally oriented to the detector. 

Now consider the Enhanced LIGO detectors, which are scheduled to begin a 
science run in mid-2009~\cite{Adhikari06}. 
They are sensitive to signals 
with frequencies higher than $\approx30$\,Hz. The signal as represented 
in Figs.~\ref{fig:FreqAmp} and 
\ref{fig:waveform} could therefore be detected from roughly 
the point where the frequency sweeps through 30\,Hz; at lower frequencies
the detector noise completely swamps the signal. We see that the 
point at which the signal enters the detector band depends on the 
binary's total mass. If the mass is too high ($\apprge 600\,M_\odot$), 
the final ringdown frequency will 
be below 30\,Hz and the signal will never enter the detector's sensitivity
band. If the mass is too low, then the detector will see much more of the 
waveform than shown in Figs.~\ref{fig:FreqAmp} and \ref{fig:waveform}. 
In that case this numerical waveform will describe only {\it part} of the 
detected waveform, and will therefore not be the optimal theoretical
waveform to use in a GW search. We
see then that the usefulness of a waveform depends on its length; 
if an NR waveform has a starting frequency of $M\omega_i$ 
(a dimensionless quantity calculated from the simulation), and a detector
has a low-frequency limit of $f_0$, then the waveform 
is best suited to search for binaries
with masses $M/M_\odot \apprge 32300 \omega_i / f_0$. For the Enhanced
LIGO detector, with $f_0 \approx 30$\,Hz, we can search for binaries with
masses  $M/M_\odot \apprge 1076\, (M\omega_i)$. Note also that if the mass
is very low ($\sim 1\,M_\odot$), then the merger and ringdown will occur at 
frequencies {\it too high} to be detected, and only the inspiral will be visible.

\begin{figure}
\centering
\includegraphics[width=60mm]{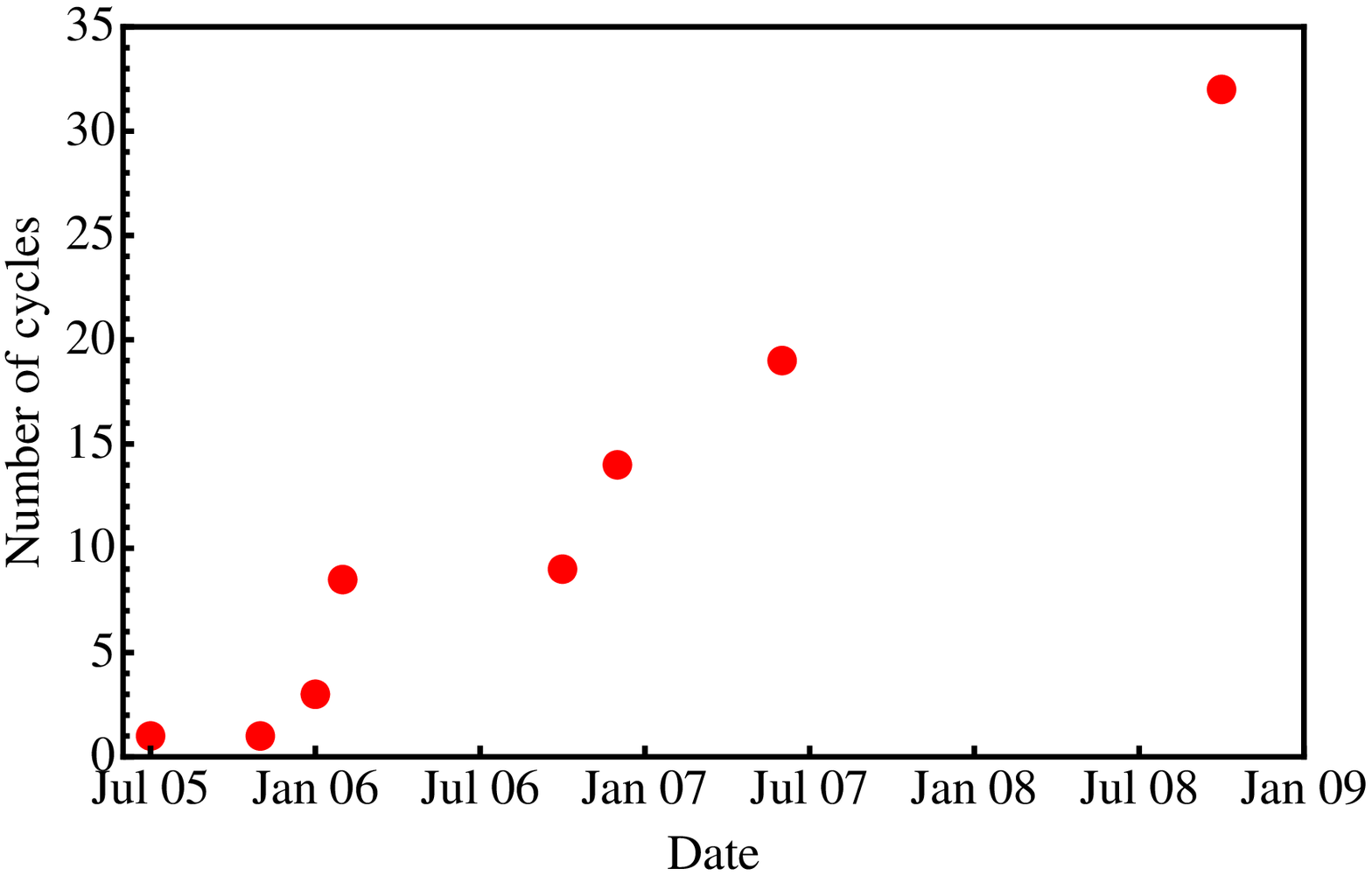}
\includegraphics[width=60mm]{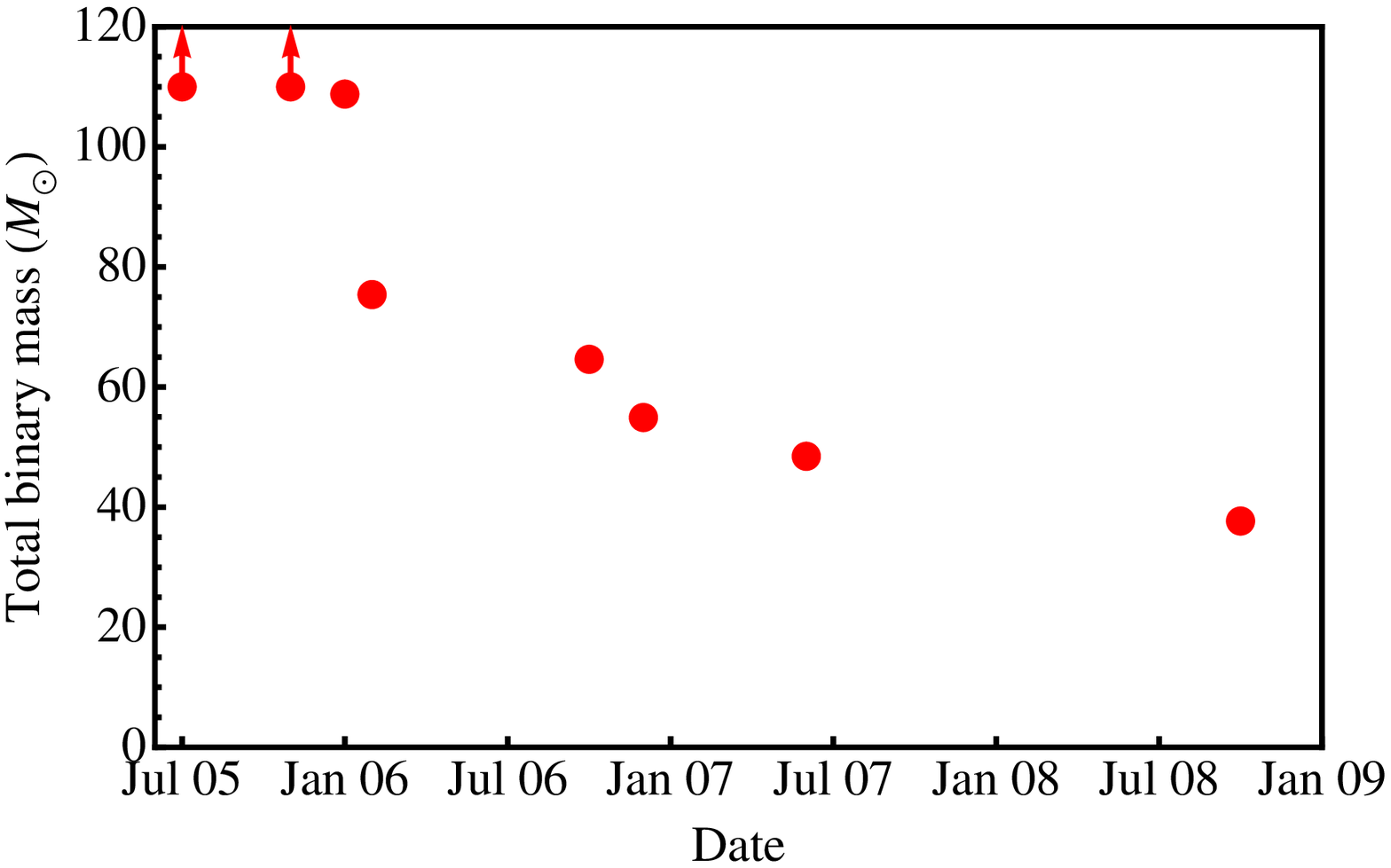}
\caption{An indication of the development of numerical simulations
  since the first simulations in 2005. The first panel shows the
  number of GW cycles before merger of an equal-mass nonspinning
  binary. The second panel shows the minimum total mass of binaries
  that could be searched for in Initial or Enhanced LIGO data with
  these waveforms; for the two points with arrows, the minimum
  detectable total mass is $M \approx 380M_\odot$. See text for
  further details.} 
\label{fig:cycles}
\end{figure}

With this discussion in mind, Fig.~\ref{fig:cycles} demonstrates the progress
of NR simulations to date, for the default case of an equal-mass nonspinning
binary following (ideally) non-eccentric inspiral.
The first published simulations in 2005 covered only half an orbit, or
one GW cycle, before  
merger~\cite{Pretorius:2005gq,Campanelli:2005dd,Baker05a}.
The first panel of Fig.~\ref{fig:cycles} shows the number of cycles
covered by successive simulations of the same
system~\cite{Campanelli:2006gf,Baker:2006yw,Buonanno:2006ui,Baker:2006ha,Hannam:2007ik,Scheel:2008rj}. The 
dramatic progress is clear: in three-and-a-half years the number of
cycles increased from one to more than thirty. What is not shown in
this plot is that the accuracy of the simulations also improved:
gravitational waves were extracted with greater precision, the
binary's eccentricity was significantly reduced, and the phase error
in the waveforms dropped by several orders of magnitude.  
(I will discuss waveform accuracy in Sec.~\ref{sec:reliability}.)
The purpose of this figure is not to suggest that progress will continue
in the same way --- in fact, it is unlikely that anyone will attempt to produce an even
longer simulation of this configuration until the appearance 
of a new generation of numerical methods --- but simply to illustrate the rate
of progress since July 2005.

A less optimistic picture from the point of view of GW detection is given in 
the second panel of Fig.~\ref{fig:cycles}, which shows the lowest-mass binary
that we could search for with these waveforms, based on the discussion above.
For the first simulations, $M\omega_i =
0.356$~\cite{Campanelli:2005dd}, and the resulting waveforms were
therefore appropriate for searches in Enhanced LIGO 
data for binaries with $M \apprge 383\,M_\odot$; the two data points
that represent these waveforms~\cite{Campanelli:2006gf,Baker:2006yw}
are off the scale of the plot. 
In three-and-a-half years the low-mass limit has improved by an order of 
magnitude to just below $40\,M_\odot$. However, we also see that the
decrease in the lower mass limit does not fall in proportion to the
increase in cycles. This is because the GW frequency increases very
slowly during the inspiral, until a few cycles before merger. If we
want a waveform that can detect 
binaries with total masses below, for example, five solar masses, we
will need waveforms that are not just eight times longer than the last
point in Fig.~\ref{fig:cycles}~\cite{Scheel:2008rj}, but hundreds of
times longer. This is a serious problem, because the most likely
sources for the first detection of GWs are predicted to have masses
lower than 50\,$M_\odot$~\cite{Flanagan:1998a}. 

Figure~\ref{fig:cycles} suggests that at some point it becomes
computationally inefficient to produce increasingly longer
simulations. How then are we to produce templates to detect low-mass
binaries? One way is to connect an NR waveform for the last orbits and
merger to a long inspiral waveform predicted by the post-Newtonian
(PN) approximation --- {\it if} the PN waveform is accurate enough. I
discuss the validation of PN waveforms by comparison with full GR
numerical waveforms in Sec.~\ref{sec:pn}, and the construction of
hybrid PN-NR waveforms in Sec.~\ref{sec:complete}. Hybrid waveforms
can in principle be used to detect binaries of any mass, within the
limits of a given detector. 

This does not solve the problem of producing all the waveforms necessary
to search for GWs from black-hole binaries, because we have considered 
only one configuration: an equal-mass
nonspinning binary with (ideally) zero eccentricity. The parameter
space of black-hole binaries is far larger. It includes the mass ratio
$q = M_1/M_2$ of the two black holes, their spins $\mathbf{S}_1$ and
$\mathbf{S}_2$, and the binary's eccentricity, $e$. 

Table~\ref{tab:LongWaveforms} summarizes the sections of this parameter 
space that have been covered by simulations to date. Keeping in mind the 
ultimate goal of producing hybrid waveforms (or other forms of complete
waveform; see Sec.~\ref{sec:complete}),
I focus only on simulations that
include more than ten GW cycles (about five binary orbits) before merger. 
This summary should serve as a quick reference of the status of 
simulations useful for GW detection. It may turn out that shorter
waveforms are sufficient to produce detection templates, or conversely
that they need to be much longer. The choice of ten orbits seemed a
reasonable limit based on PN-NR comparisons to date (see for example
\cite{Hannam:2007ik,Hannam:2007wf}), and also manages fairly
successfully to separate out those simulations that were performed
with GW observation applications as the primary motivation.

\begin{table*}
\caption{
\label{tab:LongWaveforms}
Black-hole-binary configurations for which numerical waveforms exist, and which
include at least ten GW cycles before merger.}
\begin{tabular}{|l|c|c|l|}
\hline
Configuration & Cases & Cycles & Reference \\
\hline
{\bf Equal mass} & & & \\
Nonspinning &   $e \sim 0$   & Up to 32 &  \cite{Baker:2006ha,Hannam:2007ik,Scheel:2008rj}.\\
Nonspinning & $e \sim 0.1,0.2,0.3$ & 12 -- 21 & \cite{Hinder:2008kv,Sperhake:2007gu} \\
Equal parallel spins & $|a/m| \leq 0.925$ & Up to 20 & \cite{Hannam:2007wf,Dain:2008ck} \\
\hline
{\bf Unequal mass} & & & \\
Nonspinning ($q = M_1/M_2$) & $q \leq 6$ & $12-20$ &  \cite{Buonanno:2007pf,Ajith:2007kx,Baker:2008mj} \\
Precessing spins & $a/m \sim (0.6, 0.4)$ & $\approx 20$ & \cite{Campanelli:2008nk} \\
\hline
\end{tabular}
\end{table*}

Having provided this broad overview, I will now turn to the details of 
producing NR waveforms, and current efforts to make them useful for
GW observations. Section~\ref{sec:methods} summarizes what I consider to be the
most relevant details of current numerical methods: computational
issues, black-hole-binary initial data, evolution systems, and gravitational-wave 
extraction. More details on these topics can be found in~\cite{Alcubierre2008,Pretorius:2007nq}. 
Section~\ref{sec:accuracy} addresses the physical and numerical accuracy of
simulations and the consistency between results from different codes. 
With the production of complete waveforms in mind, Sec.~\ref{sec:pn} discusses
work on comparing PN results with full GR. Section~\ref{sec:complete} is
devoted to methods to construct complete waveforms and families of waveforms 
for template banks. In Sec.~\ref{sec:issues} I highlight a number of
issues for future numerical simulations and their use for GW observations.

\section{Production of numerical waveforms}
\label{sec:methods}

In this section I will briefly summarize current techniques to
numerically solve Einstein's equations for a black-hole-binary
spacetime.

\subsection{Length scales and computational resources} 

The defining length scale of a simulation is some measure of the total
mass $M$ of the physical system. This is either chosen as the sum of
the black-hole masses, $M = M_1 + M_2$, or the total
(Arnowitt-Deser-Misner, or ADM) energy of the spacetime, $M =
M_{ADM}$. The difference between the two measures is small, and is
often referred to as the binding energy of the binary. The length
scale $M$ is then used to characterize all measures in a simulation; a
simulation may for example consider two black holes each with mass
$M/2$, initially $10M$ apart, which evolve for about $1000M$ before
merging, and emitting gravitational waves that are calculated $100M$
from the binary's center of mass. This overall scaling with $M$ means
that we can use one simulation to describe the waveform from a binary
of any mass.  

To accurately simulate a black-hole-binary spacetime, a computer code
must adequately resolve both the region near the black hole, and the
spacetime far away, where gravitational waves are extracted. In the
codes used to produce the results I will discuss here, resolutions of
at least about $M/16$ are needed near the black holes. By contrast,
the wavelength of the dominant mode of the gravitational waveform is
on the order of $10M$ during the merger phase, and so resolutions on
the order of $\sim M$ are needed in the region where gravitational
waves are extracted, which can be as far as several hundred $M$ from
the binary. To deal with such large differences in resolution
requirements on different parts of the computational domain many codes
use mesh refinement methods \cite{Berger84}. Another technique is to
use a coordinate transformation that changes the effective resolution
in different regions; such a ``fisheye'' transformation was used in
early results from the {\tt LazEv} code
\cite{Campanelli:2005dd,Campanelli:2006gf,Campanelli:2006uy,Campanelli:2006vp,Campanelli:2007ew},
and was also used in more recent simulations by the UIUC group, for
example \cite{Etienne:2007hr}. A third option is to divide the
computational domain into a number of different domains or patches,
and use a different numerical resolution and even different coordinate
systems in each domain; a multi-domain method is used in the {\tt
  SpEC} code \cite{Scheel:2006gg}. The problem of length scales
becomes even greater when we consider the sub-dominant modes of the GW
signal, which are both weaker and have wavelengths much smaller than
the dominant mode; for example, if we consider modes up to $\ell=6$,
then the ($\ell=6,m=6$) mode will have a wavelengh one third of the
length of the dominant $(\ell=2,m=2)$ mode, and the amplitude may be
several orders of magnitude smaller. 

Ideally the outer boundary of the computational domain is located at
spatial or null infinity. The only long-term binary evolution code
where one of these techniques is employed is that of Pretorius, where
spatially compactified coordinates are
used~\cite{Pretorius:2004jg,Pretorius:2005gq}. The region near the 
outer boundary is by definition poorly resolved, but a filtered buffer
zone between the well- and poorly-resolved regions is used to reduce
the build-up and propagation of any resulting errors. In all other
codes the outer boundary of the computational domain is not at spatial
infinity, and boundary conditions must be imposed. The physically
correct outer boundary conditions are not known for a black-hole-binary
spacetime, and approximate boundary conditions must be used. The BSSN
codes generally use Sommerfeld-like outer boundary conditions (which
are physically correct only for a spherically symmetric wave pulse on
a flat background), and the outer boundary is placed as far from the
binary system as computational resources allow, typically on the order
of $\sim1000M$. The effect of the outer boundary errors on the predicted
GW signal (extracted at $\sim100M$) is small, and is usually estimated
as comparable or smaller than other error sources in the code. The
Caltech-Cornell {\tt SpEC} code uses a set of constraint-preserving
boundary conditions~\cite{Rinne:2007ui} that provide a far better
approximation to the correct physics of outgoing waves on a dynamical
spacetime than Sommerfeld conditions, and make it possible to place
the outer boundary at $<1000M$ and still achieve accurate
results~\cite{Boyle:2007ft}. 

These simulations require large computational resources. Long
black-hole-binary simulations are typically run on multiple processors
of a supercomputer, and we can get an impression of the ``size'' of a
simulation from the amount of memory it requires, and the number of
CPU hours it takes to run. I did not include computational costs in
Tab.~\ref{tab:LongWaveforms} because they are not always published. As
an example, however, the highest-accuracy equal-mass nonspinning
waveform produced in~\cite{Hannam:2007ik} took roughly 18 days running
on 24 processors, for a total of about 10,000 CPU hours. 

The reader interested in the technical details of the codes currently
in use, and how they differ, is referred to Section~2
of~\cite{Aylott:2009ya} as a useful starting point.

\subsection{Initial data} 

Astrophysical black holes ultimately form through gravitational
collapse of matter, but in a black-hole simulation one need not
describe the matter at all. The black hole can instead be represented
purely through its effect on the spacetime geometry. The spacetime
singularity at the center of a black hole is difficult to describe
numerically, and there are two approaches to this problem. One is to
terminate the computational domain before it reaches a singularity;
this is called ``excision'' of the black hole \cite{Thornburg87}. No
information can escape a black hole, and so the rest of the spacetime
is unaware that the black-hole interior is missing from the numerical
solution. In practice excision is not as simple as it sounds: we  
must specify appropriate boundary conditions on the excision surface,
and we must ensure that the numerical representation of the Einstein
equations respect the speed-of-light limit on information propagation;
although physical information cannot escape the black-hole,
non-physical numerical or gauge information can in principle escape,
and may lead to numerical instabilities. Excision is used in
Pretorius's
code~\cite{Pretorius:2004jg,Pretorius:2005gq,Pretorius:2006tp}, and in
the {\tt SpEC} code~\cite{Scheel:2006gg}.  

Pretorius's original simulations began with scalar-field initial data,
chosen such that it would quickly collapse to form a black hole. Once
the black hole had formed, the interior (and the remaining scalar
field) were excised. Some of his later simulations, as well as those
performed with the {\tt SpEC} code, used excision data generated by
solving the conformal-thin-sandwich
formulation~\cite{York99,Pfeiffer:2002iy} of the initial-value
equations of general relativity, subject to inner boundary conditions
that lead to either co-rotating or irrotational black holes, and outer
boundary conditions that effectively specified the orbital speed of
the binary~\cite{Cook:2001wi,Cook:2004kt,Caudill:2006hw}. 

The second method of avoiding singularities is to choose coordinates
that bypass them: the black holes are initially described with
topological wormholes, such that as the numerical 
coordinates approach one of the black holes, they pass through a
wormhole and instead of getting closer to the singularity end up
further away, in a new asymptotically flat region. A coordinate
transformation is performed to compactify these wormholes, and the
extra asymptotically flat regions are reduced to single points, called
punctures \cite{Beig94,Beig:1994rp,Dain01a,Brandt97b}. Alternatively
one may choose ``trumpet'' coordinates such that as we approach the
black hole, we find that we are getting no closer to the singularity,
but are instead following an infinitely long cylinder
\cite{Hannam:2008sg}. These cylinders, or trumpets, can also be
compactified to punctures, and in fact this is the representation of
the black holes that simulations that start with wormhole-puncture
initial data ``naturally'' evolve towards
\cite{Hannam:2006vv,Hannam:2006xw,Brown:2007tb,Hannam:2008sg}.  

All of the initial data used to produce the waveforms I will discuss
are conformally flat, meaning that the spatial metric on the initial
slice $\gamma_{ij}$ is related to the flat metric $\delta_{ij}$ by a
conformal factor $\psi$ as $\gamma_{ij} = \psi^4 \delta_{ij}$.  
A spacetime that contains an orbiting black-hole binary will not be
conformally flat, even if we neglect the gravitational-wave
content. The use of conformal flatness simplifies the construction of
the initial data, but leads to a burst of junk radiation as 
the black holes settle down to ``true'' boosted Schwarzschild or Kerr
black holes. The junk radiation quickly leaves the system, and does
not seem to adversely affect the physics. However, the use of
conformal flatness places an effective limit on the spin of each black
hole of $a/m \apprle
0.93$~\cite{Dain:2002ee}, and the  
junk radiation severely limits the accuracy of simulations of black
holes boosted to highly relativistic speeds
\cite{Sperhake:2008ga,Shibata:2008rq}. Approaches to move beyond
conformal flatness have been proposed for both 
excision~\cite{Matzner98a,Marronetti00a,Marronetti00,Bonning:2003im,
Lovelace:2008tw,Lovelace:2008hd,Cook:2008mv} 
and puncture data~\cite{Hannam:2006zt,Shibata:2008rq,Kelly:2007uc}, although 
they have not yet been used for the long-term orbit simulations 
of the kind we are focussing on here. 

\subsection{Evolution systems} 

Given black-hole-binary initial data, a stable evolution requires a
numerically well-posed and stable formulation of Einstein's equations,
as well as a judicious choice of gauge conditions. Finding a suitable
set of evolution equations and gauge conditions was one of the major
problems in the field during the decade preceeding the 2005
breakthroughs. The textbook~\cite{Alcubierre2008} contains a review of
this topic, and the mathematical issues are discussed further
in~\cite{Friedrich:2000qv,Husa:2008jx}; and one illustration of the
severity of the problems is the Apples with Apples
project~\cite{Alcubierre:2003pc,Babiuc:2007vr}.  

Although not all mathematical and numerical questions have been
resolved, long-term stable simulations can now be performed with
either a variant of the generalized harmonic
formulation~\cite{Friedrich85,Friedrich:2000qv,Pretorius:2004jg,Lindblom:2005qh} 
or the moving-puncture treatment~\cite{Campanelli:2005dd,Baker05a} of
the Baumgarte-Shapiro-Shibata-Nakamura
(BSSN)~\cite{Shibata95,Baumgarte99} formulation. A review of
successful methods and applications for black-hole-binary simulations
is given in~\cite{Pretorius:2007nq}.

The generalized harmonic formulation deals directly with the full spacetime
metric, $g_{\mu\nu}$. The metric evolves via a set of generalized wave 
equations, which are in a manifestly hyperbolic form. In standard harmonic
coordinates, the spacetime coordinates $x^\mu$ are chosen to satisfy
$\Box x^\mu = 0$. This condition is now generalized so that the gauge
(coordinate) conditions are specified by source functions 
$\Box x^\mu = H^{\mu}$~\cite{Friedrich85,Friedrich:2000qv,Pretorius:2004jg},
which in turn are either specified functions of time, or satisfy their own
evolution 
equations~\cite{Pretorius:2005gq,Pretorius:2006tp,Boyle:2007ft,Szilagyi:2006qy,Scheel:2008rj}.

The BSSN decomposition starts instead with the (numerically ill-posed)
ADM-York equations for the spatial quantities
$(\gamma_{ij},K_{ij})$~\cite{Arnowitt1962,York79}. The BSSN
reformulation provides evolution equations for conformally rescaled
quantities,
$\{\psi,K,\tilde{\gamma}_{ij},\tilde{A}_{ij},\tilde{\Gamma}^i\}$,
where  $\gamma_{ij} = \psi^4 \tilde{\gamma}_{ij}$ and $K_{ij} = \psi^4
(\tilde{A}_{ij} + \tilde{\gamma}_{ij} K)$, and the extra variable,
$\tilde{\Gamma}^i = \partial_j \tilde{\gamma}^{ij}$ is introduced. 
The moving-puncture extension of the BSSN system deals with puncture 
data, and involves introducing either $\phi = \ln \psi$~\cite{Baker05a}, 
$\chi = \psi^{-4}$~\cite{Campanelli:2005dd} or $W =
\psi^{-2}$~\cite{Marronetti:2007wz}, and evolving that quantity   
instead of the conformal factor $\psi$, and specifying
gauge conditions that allow the punctures to move across the numerical
grid. Although developed heuristically over time, these gauge conditions
serendipitously attract the numerical slices to approximate stationarity
(so that the variables evolve largely due to the physics, and not mere
coordinate changes), and move the punctures across the grid. The
behaviour of these gauge conditions for moving-puncture evolutions of
a Schwarzschild black hole are discussed in detail 
in~\cite{Hannam:2008sg}. The numerical stability of 
 the moving-puncture system is considered
 in~\cite{vanMeter:2006vi,Gundlach:2006tw}. 
  
 The generalized-harmonic and moving-puncture methods have been
 found to work for simulations of up to 15 orbits, for binaries with
 significant eccentricity, with mass ratios up to 1:10, and spins up
 to the conformal-flatness limit of $a/m \sim 0.93$. Despite this
 wealth of evidence that these methods work, surprisingly little has
 been done to explain why. The properties that are known to be
 necessary for a stable simulation (in particular, a symmetric
 hyperbolic evolution system), are also known to  
 not be sufficient. What distinguishes these methods from others? Could it
 be that most other (well-posed) systems of equations can be stably 
 evolved with appropriate gauge conditions and methods to move the 
 black holes through the grid? Are there situations where the current 
 methods will fail? These questions have been largely neglected, and 
 deserve more attention.

\subsection{Gravitational-wave extraction} 
\label{sec:waves}

Finally, once we have have evolved a black-hole-binary system through its
last orbits, merger and ringdown, we want to measure the gravitational-wave
signal that was emitted. This is done by calculating either the Newman-Penrose
scalar $\Psi_4$ \cite{Newman62a,Stewart:1990uf}, which is a
measure of the outgoing transverse gravitational radiation in an
asymptotically flat spacetime, or the odd- and even-parity master functions
$Q^{+}$ and $Q^{\times}$ 
in the Zerilli-Moncrief formalism (see~\cite{Nagar:2005ea} for a review). 
The wave strain, $\hc = h_+ - \mathrm{i} h_\times$, is related to $\Psi_4$ by 
two time integrations, and to $Q^{+,\times}$ by one time integration.  
The quantities are calculated on spheres of constant
coordinate radius some distance from the binary, and then decomposed 
in spin-weight $s=-2$ spherical harmonics $_{s}Y_{\ell m}$. Not only are the 
$(\ell,m)$ harmonics more easily compared with analytic predictions of the 
GW signal, which are usually presented in the same way, but calculating the 
contribution to each mode involves an integration over the entire sphere, 
\begin{equation}
\hc_{\ell m} \equiv \langle _{-2}Y_{\ell m}, \hc \rangle = 
        \int_0^{2\pi} \dphi \int_0^{\pi}
      \hc \, _{-2}\overline{Y}_{\ell m}\, \sin \theta\,\dtheta\,
      \label{eq: scalar_product},
\end{equation} which effectively
smooths out numerical noise; for example, 
$\Psi_4$ calculated at one point on the numerical
grid is typically rather noisy, while $\Psi_4$ reconstructed from a (reasonable)
number of modes will be a clean wave signal. Given the values of the spherical 
harmonics, the corresponding wave signal can be calculated for any sky location 
and binary orientation.

Producing the strain $\hc$ from $\Psi_4$ appears at first sight to be
merely a matter of performing two time integrations of $\Psi_4$ and
appropriately choosing two integration constants. However, the
numerical data contain numerical noise, and also small deviations from
the ``correct'' signal that would be seen at infinity due to gauge
effects. These small errors can be grossly magnified after an
integration over the entire duration of the waveform, and producing
$\hc$ from $\Psi_4$ requires {\it two} such integrations. One may be
tempted to first Fourier transform $\Psi_4$ to the frequency domain,
where two time integrations can be trivially performed, but this also
involves an implicit choice of integration contants, and once again
care has to be taken to produce physically reasonable
results (see for example~\cite{Campanelli:2008nk}). 

The gauge errors in $\Psi_4$ will decay with increasing extraction
radius --- or at least they will with an appropriate gauge 
choice \cite{Lehner:2007ip}, which seems to be the case with typical
gauge choices~\cite{Hannam:2007ik,Boyle:2007ft,Scheel:2008rj}; the
good agreement between results from both the Newman-Penrose and
Zerilli-Moncrief methods provides further evidence that our gauges are
well-behaved \cite{Koppitz:2007ev}. This suggests that  we can try to
reduce gauge errors by extrapolating the waveform to an estimate of
the wave that would be observed at infinity (i.e., by a distant
detector). Unfortunately the extrapolation is rarely perfectly clean,
and introduces further numerical noise that tends to produce an even
less believable version of the strain than an integration based on the
raw data.  

Despite these difficulties, one may produce a reasonable estimate of
the strain by applying a number of ``cleaning'' procedures; see, for
example~\cite{Husa:NRDAtalk2008}. An alternative option, of course, is
to calculate $Q^{+}_{\ell m}$ and $Q^{\times}_{\ell m}$ instead, from
which the strain can be produced by only one time integration, and the
results are in general far cleaner.

\section{Reliability of numerical waveforms}
\label{sec:reliability}

Before using NR waveforms for GW applications, we need to quantify their
numerical and physical accuracy. We also need to verify that the results
produced with different sets of evolution equations, initial-data, gauge 
conditions and wave extraction methods are all consistent.

\subsection{Numerical and physical accuracy}
\label{sec:accuracy}

One of the most important checks of a numerical result is a convergence test. 
A given numerical method will be expected to converge to the continuum solution as 
some function of the numerical resolution. For example, if a method to solve a 
one-dimensional problem to find a solution $f(x)$ is 
second-order accurate in the grid spacing $\Delta x$, then the numerical 
solution $\bar{f}(x)$ will be related to the true solution $f(x)$ by \begin{equation}
f(x) = \bar{f}(x) +  a(x) \Delta x^2 + {\cal O}(\Delta x^3).
\end{equation} In this case, each time we halve the grid-spacing $\Delta x$, the
error in the solution should decrease by roughly a factor of four. If
we have an analytic solution to compare with, we can readily verify
the dependence of the error on the grid spacing, and if not we look
instead at $f_{\Delta x}(x) - f_{\Delta x/2}(x)$ for several choices
of $\Delta x$. A convergence test verifies that the code is producing
a valid solution of a system of partial difference equations, and
helps determine how much numerical resolution is required for the code
to be in the convergent regime and for the solution to be sufficiently
accurate.  

In large and complex codes like those used to simulate
black-hole-binary systems, convergence tests are rarely as
straightforward as suggested above. In a mesh-refinement-based code,
different parts of the code may have different theoretical convergence
orders, and it may be far from clear how the overall convergence
should behave. Ultimately one should be able to choose a high-enough
resolution such that the lowest-order part of the code
dominates. However, 3D numerical simulations require large 
computational resources, and successively halving the grid spacing
until the predicted convergence rate is seen is usually not
practical. Simulations are often performed at the edge (or what is
perceived to be the edge) of the convergent regime, and convergence
behaviour deemed sufficiently close to theoretical expectations is
usually considered acceptable. The definitions of ``sufficiently
close'' and ``acceptable'' are of course subjective, but in most cases
the error estimates quoted in the literature err on the side of
caution. Those who remain suspicious may be reassured by
Section~\ref{sec:consistency} below.  

The uncertainties and errors in a numerical waveform are not defined 
entirely by the convergence properties of the code. There are also
errors in the physical specification of the data. The definition of
black-hole mass (used to define the overall scale of the simulation)
is not unique, nor is the 
measurement of black-hole spins. A simulation that is supposed to
model non-eccentric inspiral will always have some remaining
eccentricity, and for eccentric binaries there is (once again) no
unique definition of eccentricity. (Zero eccentricity is relatively
easy to define; it's just difficult to
achieve~\cite{Pfeiffer:2007yz,Husa:2007rh,Boyle:2007ft,Campanelli:2008nk,Walther:2009ng}) 
Finally, and often most importantly, we want the gravitational-wave
signal as measured far from the source, in the region where the wave
can be considered as a perturbation on flat space. In practice the
waves are extracted from the numerical solution only a few hundred $M$
from the source; for a binary that consists of two solar-mass black
holes, this is equivalent to measuring the gravitational waves only
about 300\,km away, about one hundred-millionth of the distance of the
Sun to the Earth! As ridiculous as this may sound, it is a large
enough distance in general relativistic terms that the amplitude error
in a numerical waveform is only a few percent --- it should simply be
borne in mind that this few-percent error may be the largest error
ingredient in the entire waveform. A good illustration of the errors
that enter into the calculation of a numerical waveform are provided
in Table~III of \cite{Boyle:2007ft}. 

Both the numerical and physical accuracy of numerical waveforms has improved 
steadily since 2005. The first simulations were performed with a code that 
resolved each time slice with second-order-accurate finite
differences~\cite{Pretorius:2005gq}. The moving-puncture results that
followed six months later~\cite{Campanelli:2005dd,Baker05a} used
second- and fourth-order-accurate finite differences. An accurate
comparison of numerical and post-Newtonian waveforms was performed in
2007 using sixth-order
finite-differencing~\cite{Husa:2007hp,Hannam:2007ik}, and the {\tt
  LazEv} code now routinely uses eighth-order
methods~\cite{Lousto:2007rj}. The {\tt SpEC} code, which has produced
the most accurate equal-mass nonspinning binary waveform to date, uses
pseudospectral methods to describe the spatial slice, and achieves an
overall error in the phase during 30 cycles of inspiral, merger and
ringdown of less than 0.1\,radian, and an uncertainty in the amplitude
of at worst (during merger and ringdown) of
1\%~\cite{Boyle:2007ft,Scheel:2008rj}.

\subsection{Consistency between solutions}
\label{sec:consistency} 

The uncertainty estimates published with numerical waveforms lead us
to believe that those waveforms are extremely accurate. However, these
solutions were produced using different formulations of Einstein's
equations, different numerical methods and, perhaps more importantly,
different gauge conditions, families of initial data and wave
extraction methods. The effects of these different approaches,
although they are expected to be small, need to be fully quantified.  

With this goal in mind a comparison was performed between three of the
early black-hole-binary simulations in 2006~\cite{Baker:2007fb}. Since
then many much longer simulations have been performed with far greater
accuracy, and it is these more recent simulations that are likely to
play a role in future GW searches; some have already been used for the
NINJA project~\cite{Aylott:2009ya} to test GW search pipelines on NR
waveforms buried in simulated detector noise. As such, a project to
compare long waveforms in the context of GW data analysis has been
organized as a complement to the NINJA project, and has been dubbed
Samurai~\cite{Hannam:2009hh}.  

The Samurai project compares equal-mass nonspinning waveforms from
five codes, dealing with only the last $1000M$ before merger, and
$80M$ after merger. The five codes are {\tt
  BAM}~\cite{Brugmann:2008zz,Husa2007a}, {\tt
  CCATIE}~\cite{Pollney:2007ss},  
{\tt Hahndol}~\cite{Imbiriba:2004tp,vanMeter:2006vi}, {\tt MayaKranc}~\cite{Vaishnav:2007nm}
and {\tt SpEC}~\cite{Scheel-etal-2006:dual-frame}. Since the {\tt
  SpEC} waveform has the lowest associated numerical uncertainties, it
is used as the reference against which the others are compared. 

Two kinds of comparison are performed. The first deals with the phase
and amplitude agreement of the $\Psi_4$ waveforms. It is found that
all of the waveforms agree within their claimed internal
uncertainty. This acts as a clear validation of the results between
different codes, and demonstrates that the variation between waveforms
due to different numerical techniques, initial data, gauge conditions
and wave-extraction methods is at worst no larger than the internal
error estimates in each waveform.  

The details of the uncertainties in each numerical code are of little direct
interest in the practical business of GW detection and source parameter estimation. 
For that reason a second comparison attempts to assess what the apparently small
differences between the five waveforms mean for GW searches. The relevant 
quantity to compare for GW detection is the match ${\cal M}$  between two
waveforms, which quantifies their disagreement with respect to the noise 
spectrum of a given detector~\cite{Owen_B:96}. In GW searches, if the match between
the correct physical waveform and the template is greater that 0.965, then no more
than about 10\% of signals will be lost. 

\begin{figure}[tp]
\centering
\includegraphics[width=62.5mm]{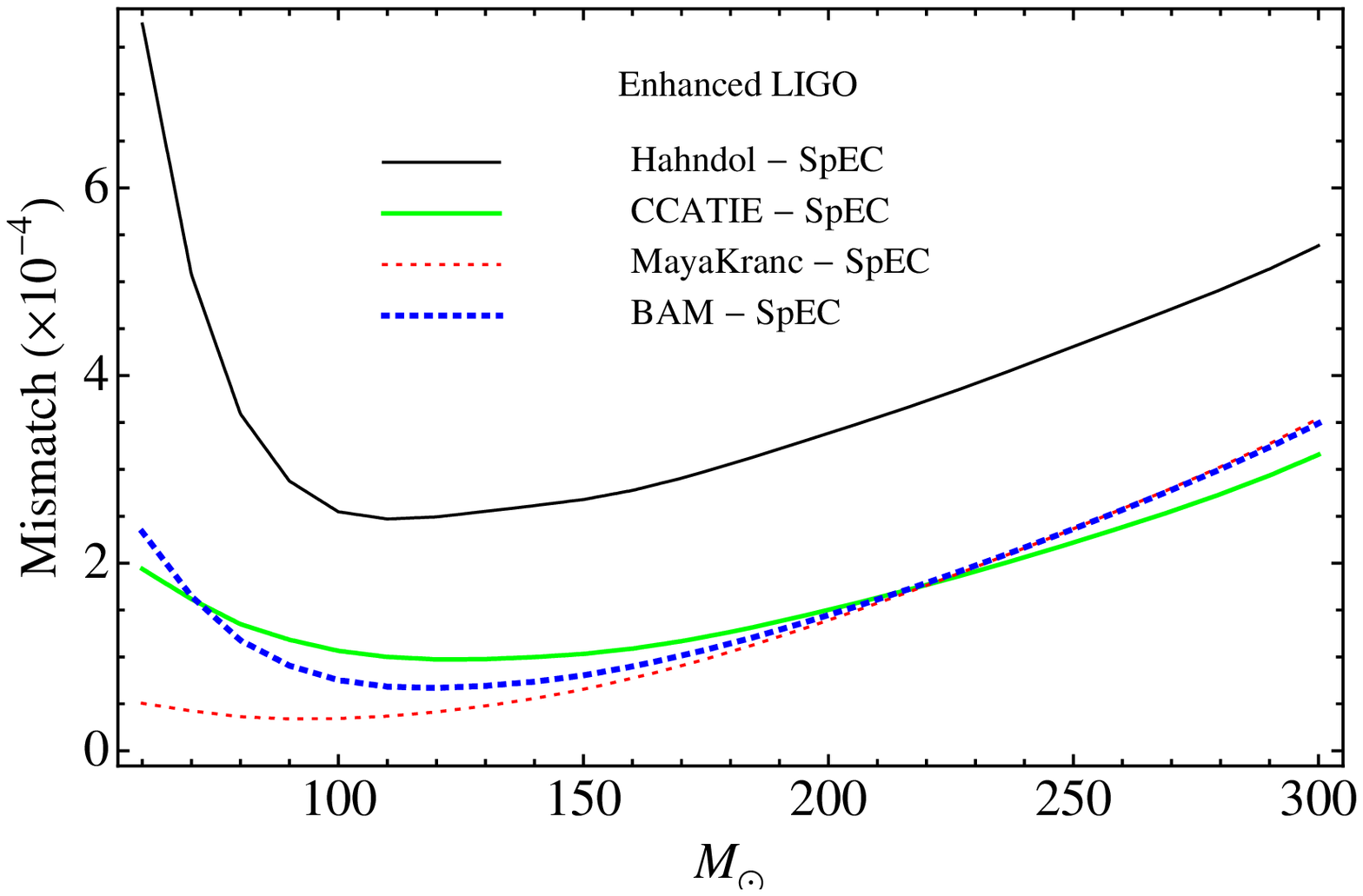}
\includegraphics[width=60mm]{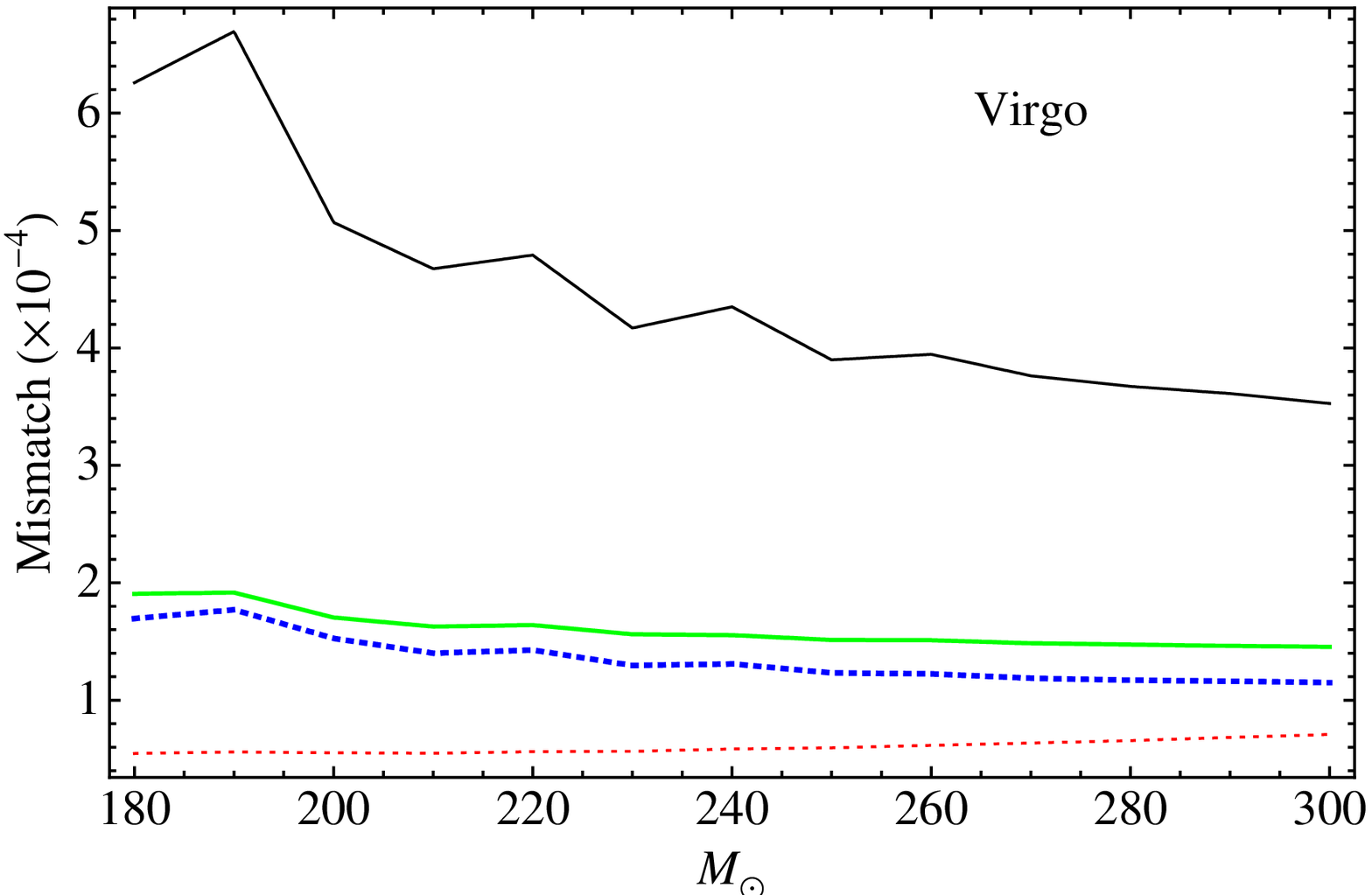}
\caption{The mismatch between the {\tt SpEC} waveform and each of the
  other codes. The two plots show the results for the Enhanced LIGO and
  Virgo noise curves. The lower end of the mass range was chosen
  such that the entire numerical waveform was included in the
  detector's frequency band. (Plot taken from~\cite{Hannam:2009hh}.)}
\label{fig:matches}
\end{figure}

For the Samurai waveforms, the matches are so close to unity that it makes more
sense to consider the mismatch, $1 - {\cal M}$. Figure~\ref{fig:matches} shows
the mismatch between the {\tt SpEC} waveform and the others for the Enhanced LIGO
and Virgo detectors. (The lower mass cut-off for each plot is
determined by the calculation  
discussed in the Introduction; the low-frequency limit for Virgo is taken as 10\,Hz.)
As is clear from the figure,
the matches are well within the standard $1 - {\cal M} < 0.035$ threshold for 
detection! 

One can also make a comparison relevant to parameter estimation. As 
discussed in \cite{Lindblom:2008cm}, if the signal-to-noise ratio of the difference
between two waveforms $\delta \hc(t) = \hc_1(t) - \hc_2(t)$ is less than one, 
then the two waveforms will be indistinguishable in a GW search. 
The SNR of both the waveforms and their difference decreases in inverse 
proportion to the distance $D$ of the detector from the source. We can
therefore determine the maximum SNR such that, if a signal were detected
with a lower SNR, it would not be possible to distinguish whether it was
$\hc_1$ or $\hc_2$. For the Enhanced LIGO and Virgo 
detectors, a detection will be considered reliable if the SNR is above
5--8, and SNRs above 30 are considered unlikely. (For example, the 
maximum SNR of injections for the NINJA project is 30~\cite{Aylott:2009ya}.) 

\begin{figure}[tp]
\centering
\includegraphics[width=60mm]{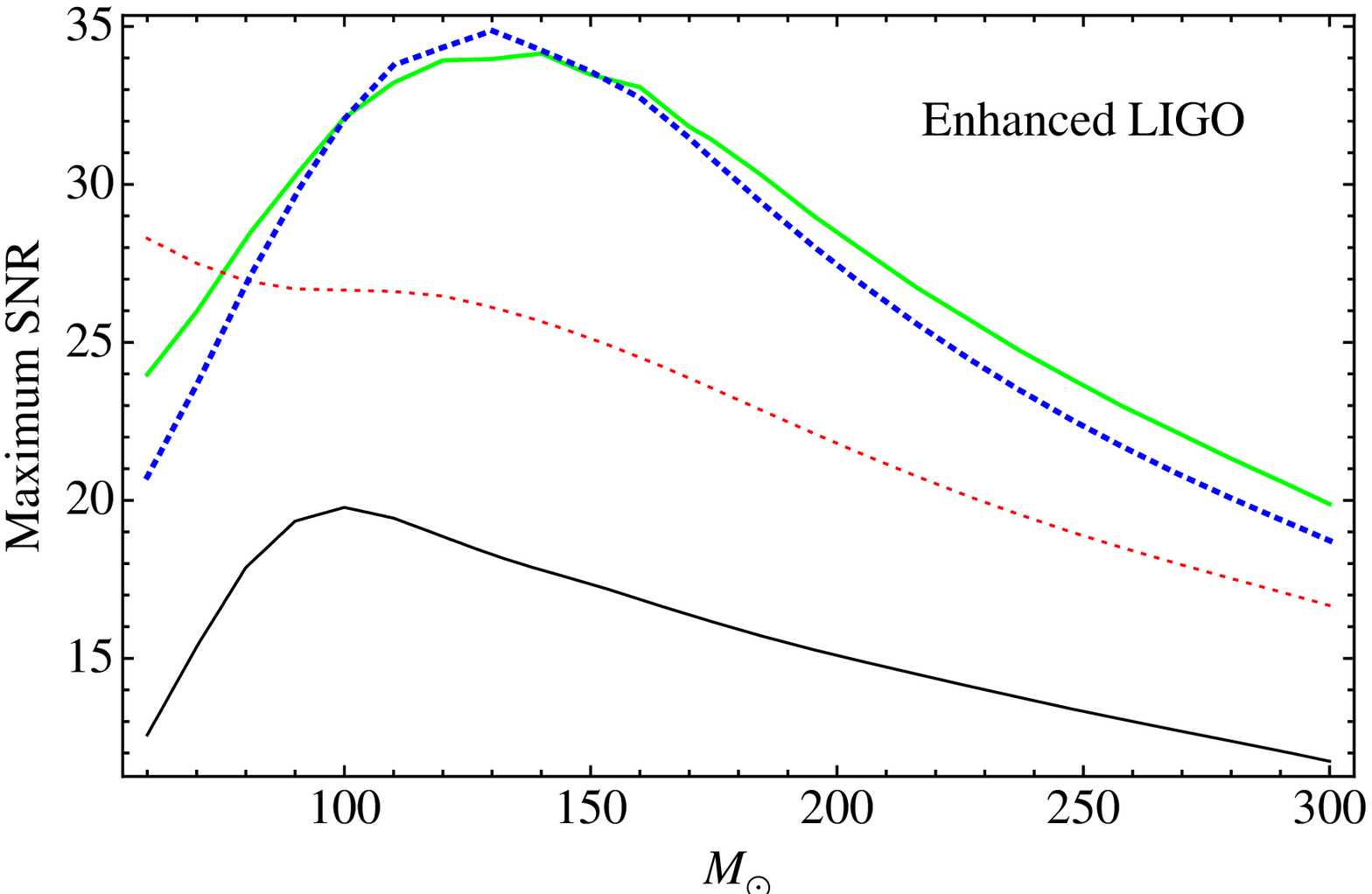}
\includegraphics[width=62.5mm]{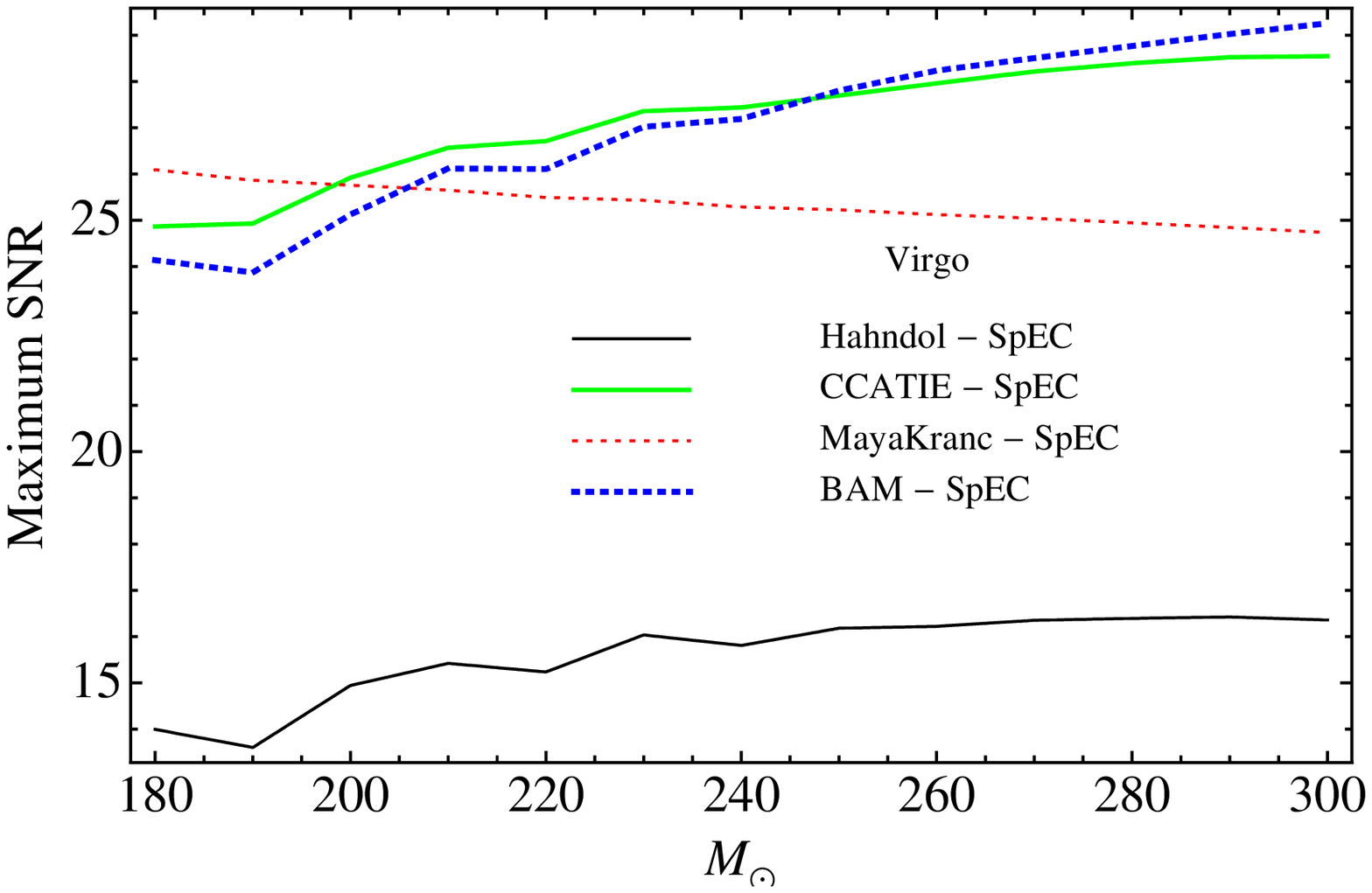}
\caption{The signal-to-noise ratio (SNR) for which the {\tt SpEC} and
  each other waveform will be indistinguishable in any measurement of
  intrinsic parameters. Results are shown for the Enhanced LIGO and
  Virgo detectors. See text for further explanation. 
  (Plot taken from~\cite{Hannam:2009hh}.)}
\label{fig:measurement}
\end{figure}

Figure~\ref{fig:measurement} shows the maximum SNR for
indistinguishability for the Samurai waveforms. The measurement
criteria was minimized over a time- and phase-shift, so these results
do not apply to a measurement of the time of arrival or initial phase
of the waveform, or any parameter that is affected by them (for
example the sky location), but it {\it does} apply to any of the
intrinsic parameters of the binary
$\{M,q,\mathbf{S}_1,\mathbf{S}_2,e\}$ that we first discussed in
Sec.~\ref{sec:intro}; this simple analysis is also limited to
single-detector searches. Within these caveats, and within the range
of binary masses shown in Fig.~\ref{fig:measurement}, we see that in
general the signals are indistinguishable for an SNR below about 25,
and in all cases cannot be distinguished if the SNR is below
$\sim14$. This suggests that these five waveforms are unlikely to be
distinguishable (for intrinsic parameters) in single detectors prior
to the commissioning of Advanced LIGO and Virgo in 2014. A
comparable study remains to be done for unequal-mass binaries with
precessing spins, which are computationally more challenging to
simulate, but recent first long simulations of such
systems~\cite{Campanelli:2008nk} suggest that similar levels of
accuracy are possible. As such we expect that the current level of
accuracy and consistency of numerical waveforms is adequate
for at least the next five years.

Beyond that time, it is important to bear in mind that these results refer 
to only the dominant mode of the GW signal. The subdominant modes 
are unlikely to change the results for GW detection (although this depends 
on the orientation of the binary), but can be important for parameter 
estimation~\cite{Sintes:1999cg,VanDenBroeck:2006ar}. Since the 
Samurai results suggest that the black-hole dynamics are
captured accurately in current simulations, achieving comparable
accuracy of the higher modes depends only on the accuracy of
the wave extraction (the distance of extraction from the source,
and the numerical resolution in the wave-extraction zone), and suggest
that this is where efforts should be directed in improving numerical
simulations; I will say more about this in Sec.~\ref{sec:issues}.

Having concluded that numerical waveforms are in general sufficiently 
accurate for use in data-analysis applications, we move on to the question
of providing waveforms of arbitrary length.

\section{Comparison with post-Newtonian predictions}
\label{sec:pn}

As we saw in the Introduction, current waveforms can only be used to
search for binaries with total masses above $\sim40\,M_\odot$, and
waveforms suitable for searches of binaries with masses below
5\,$M_\odot$ would need to be hundreds or thousands of times
longer. (References to the binary's mass will always be to the {\emph
  total} mass of the binary, $M = M_1 + M_2$.)

Fortunately we do not need to simulate all of those orbits in full general 
relativity. The wave signal from the slow inspiral can also be modeled
by post-Newtonian 
(PN) methods, and one would hope that the PN approximation is adequate up 
to the point where we begin our numerical simulations, and that it is
possible to smoothly 
connect the PN and NR signals to produce a ``complete''
waveform. I will discuss work on producing such complete waveforms in 
Section~\ref{sec:complete}. But first one must quantify the level of accuracy of the
PN approximants. 

\begin{figure}
\centering
\includegraphics[angle=0,clip,width=.45\textwidth]{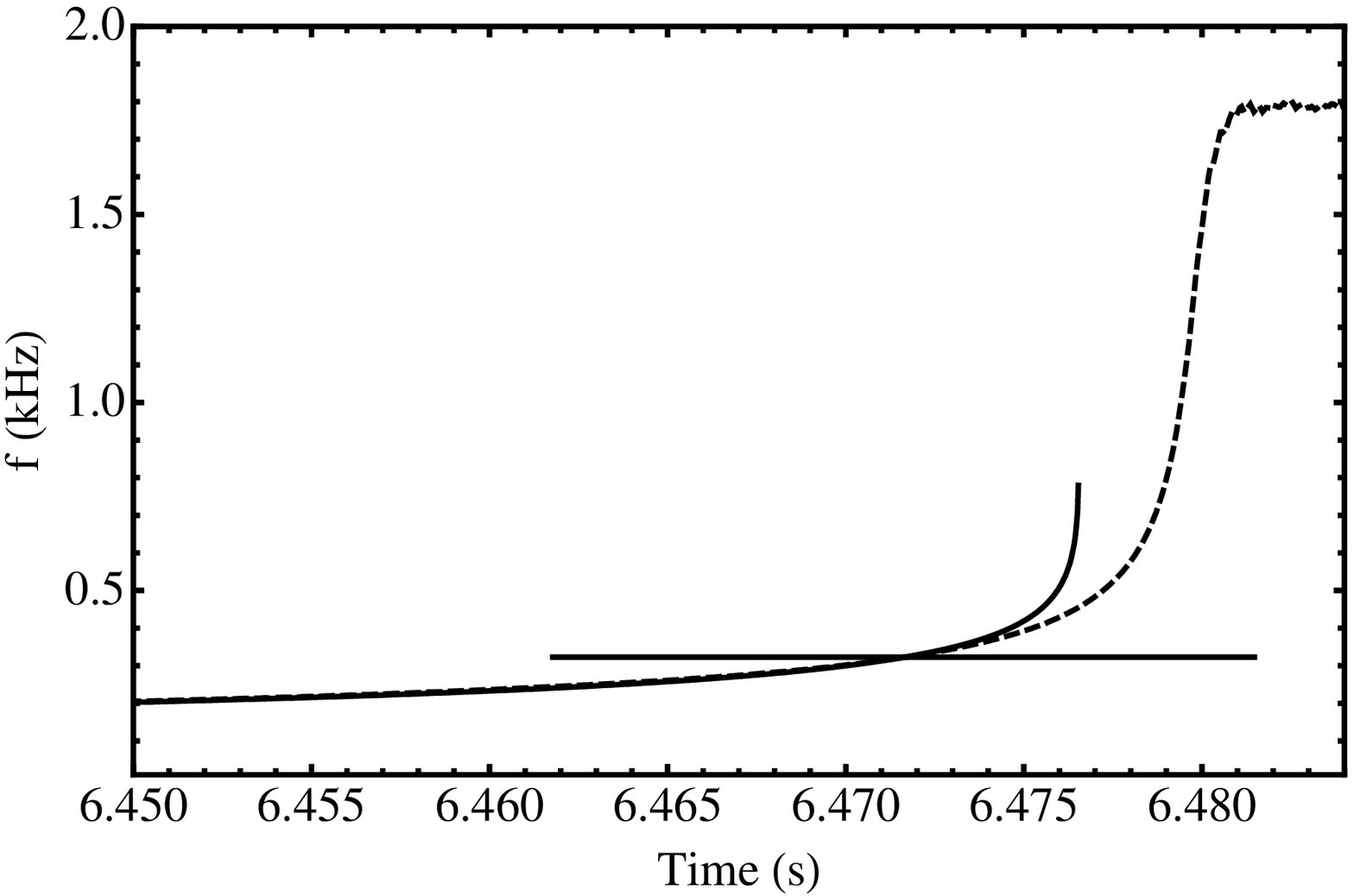}
\includegraphics[angle=0,clip,width=.445\textwidth]{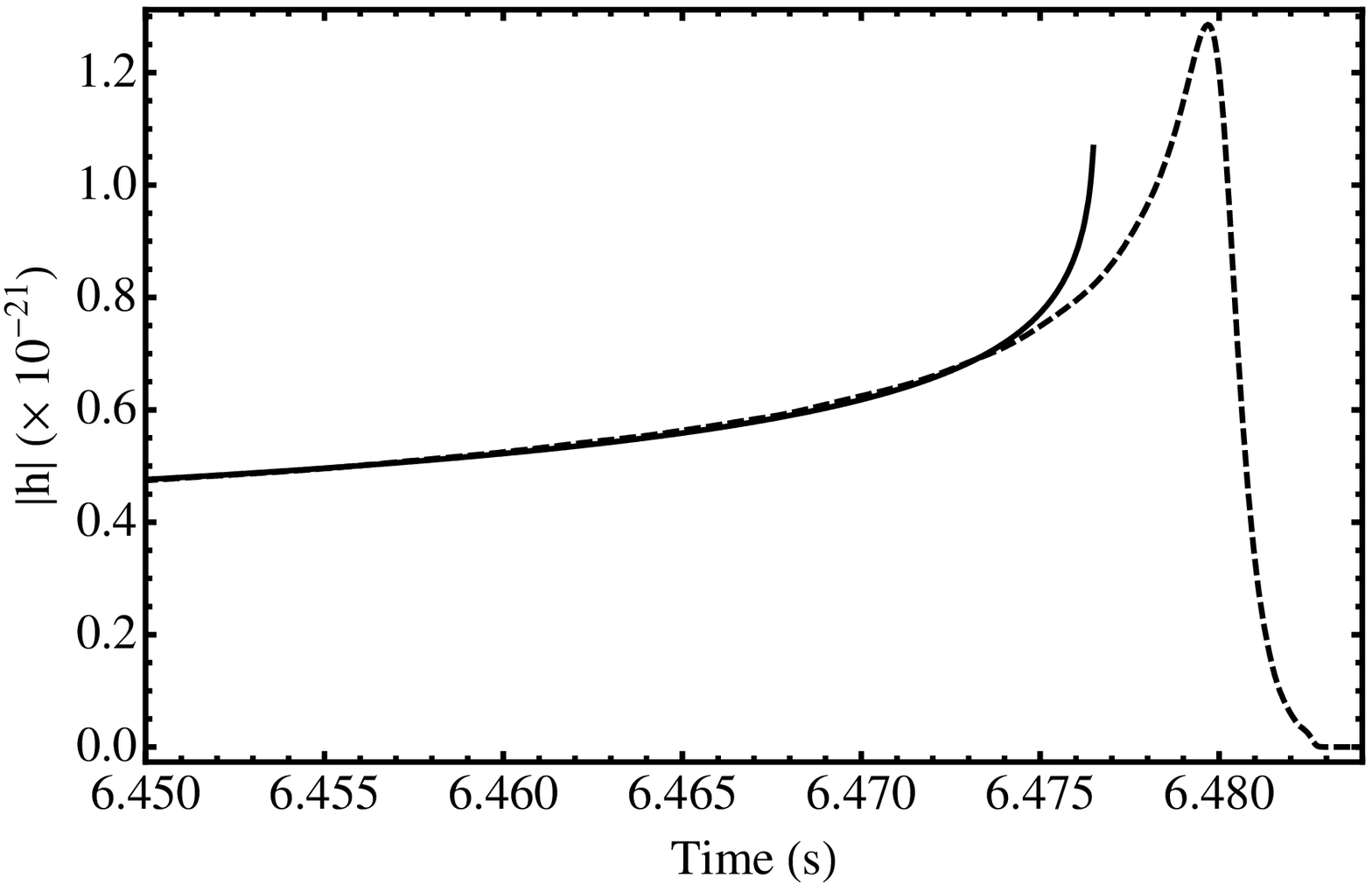}
\caption{GW frequency and amplitude evolution for a 10\,$M_\odot$ binary starting
about four orbits (eight cycles) before merger. The solid line indicates the PN values 
(using TaylorT1 phase and quadrupole amplitude) and
the dashed line shows the NR values. The PN result is cut off just before it diverges.
The results were time-shifted so that the frequencies agree at $M\omega = 0.1$, 
indicated by the horizontal line in the first panel.}
\label{fig:10Mexample}
\end{figure}

Consider a 10\,$M_\odot$ equal-mass nonspinning binary. PN
calculations tell us that the Enhanced LIGO sensitivity band will
contain about 150 orbits (300 cycles) before merger, and the signal
will last about 6.5\,s. The first panel of Fig.~\ref{fig:10Mexample}
shows the frequency evolution after 6.45\,s. Although the plot
includes only 0.05\,s of the 6.5-second-long signal, this accounts for
arond 20\% of the power output of the entire signal; see Fig.~4
in~\cite{Pan:2007nw}. The solid line shows the PN frequency as a
function of time, and the dashed line shows the NR result. The PN line
was produced with the TaylorT1 approximant, and was cut off just
before it diverges. Similar results would be obtained with any other
standard (i.e., Taylor-expanded) PN approximant; the figure was
produced purely as an illustration of the general behaviour of PN and
NR results. The second panel of Fig.~\ref{fig:10Mexample} shows the
corresponding time development of the amplitude of the $(\ell = 2,
m=2)$  mode of the GW strain if the binary were optimally oriented to
the detector and located 100\,Mpc away. The PN and NR waveforms were
aligned in time such that $M\omega = 0.1$ at the same time for each.

The PN and NR phase and amplitude appear to agree well until just before the PN 
approximant diverges. The horizontal line in the frequency plot indicates the point at which 
$M\omega = 0.1$; it seems reasonable to conclude that PN and NR agree fairly well up 
to that point. This demonstration of the qualitative agreement between NR and PN 
phase and amplitude mimics that performed with more care in the first published
NR-PN comparison in~\cite{Buonanno:2006ui}. In that work the leading PN 
(i.e., quadrupole) amplitude and 3.5PN TaylorT3 phase were found to agree 
reasonably well with the NR quantities up to about one quarter of an orbit before 
merger. Having observed this good qualitative agreement, we now wish to make 
the comparison more precise.

Detailed comparisons between PN and NR waveforms for equal-mass nonspinning
low-eccentricity binaries have been reported 
in~\cite{Baker:2006ha,Hannam:2007ik,Gopakumar:2007jz,Boyle:2007ft}. 
It was found that most 
PN approximants (at 3.5PN order) predict the phase to within about a radian of
the full GR result for the last 14-25 cycles before $M\omega = 0.1$, but that
one approximant, TaylorT4, predicts the phase to within 0.05\,rad, although this
agreement is assumed to be accidental~\cite{Boyle:2007ft}. The leading 
quadrupole amplitude is found to disagree with NR results by about 
6\%, the 2.5PN amplitude disagrees by about 2\%, and the 3PN amplitude
disagrees by less than 1\%~\cite{Hannam:2007ik,Boyle:2007ft}.

NR-PN comparisons moved beyond the equal-mass nonspinning case
in~\cite{Hannam:2007wf}, where 
spinning binaries were studied. The black holes had equal mass and equal spin, 
with the spin parallel to  the orbital angular momentum of the binary. The spins 
considered were 
$a/m \sim \{0,0.25,0.5,0.75,0.85\}$. Phase comparisons were performed against the TaylorT1,
T4 and Et approximants. The impressive phase accuracy of the T4 approximant was 
found to quickly deteriorate in the spinning case, and for black holes with spins of $a/m\sim0.85$,
the phase disagreement over the ten cycles up to $M\omega=0.1$ was $\sim2$\,rad. 
The TaylorT1 approximant performed most consistently for all values of spin, and the
largest phase disagreement was less than 1.5\,rad in the highest-spin case. 

When the spins are aligned parallel (or anti-parallel) to the orbital angular momentum,
they remain constant throughout the evolution. In other cases the spins precess. 
A binary with precessing spins was studied in~\cite{Campanelli:2008nk}; the mass 
ratio was $q = 0.8$ and
the spins were $a_1 \sim 0.6$ and $a_2 \sim 0.4$. The authors compared their numerical
results with PN waveforms constructed from the orbital motion predicted by an integration
of the PN equations of motion, and found that the phase agreed within about 4\,rad over
$\sim10$ cycles before merger, for the $(\ell=2,m=2)$ mode, although the details of
the comparison make it difficult to directly compare with the published equal-mass
PN-NR comparisons. 

The example of the TaylorT4 approximant illustrates that the performance of a given
PN approach may vary dramatically between black-hole-binary configurations. 
One way to test the robustness of a PN method is to examine not the final prediction
of the GW phase and amplitude, but the ingredients that go into the calculation, 
the energy flux $F$ and derivative of the center-of-mass energy $dE/d\omega$.
These were studied for equal-mass nonspinning binaries in~\cite{Boyle:2008ge}.
The authors found that none of the Taylor PN 
approximants predicted a flux that agreed well with NR results over the last
25 cycles before merger from an equal-mass nonspinning binary,
providing further evidence that the good agreement of various PN approximants
for differing black-hole-binary configurations is indeed accidental (see Fig.~9 in
\cite{Boyle:2008ge}). In contrast, effective-one-body (EOB) and Pad\'e-resummed 
approximants accurately predicted both $F$ and 
$dE/d\omega$, suggesting that although these approaches do not predict the
phase in the equal-mass nonspinning case as well as the TaylorT4 approximant,
the EOB and Pad\'e methods may be more robust when applied to other 
configurations. This prediction remains to be tested. 

For the configurations for which NR and PN results have been compared, however,
all of the standard approximants appear to perform adequately enough that we can
move on to the question of connecting them to produce complete waveforms, 
which I will discuss in the next section.

\section{Complete waveforms for GW searches}
\label{sec:complete}

One way to  produce complete waveforms
that cover an arbitrary portion of the inspiral, plus the merger and ringdown, is to 
connect NR and PN waveforms to produce hybrid waveforms. A method to do this 
was first suggested in \cite{Pan:2007nw} and later 
in~\cite{Ajith:2007qp,Ajith:2007kx,Ajith:2007xh}. 

In \cite{Pan:2007nw} PN waveforms computed using a 3.5PN phase and 
quadrupole amplitude are matched with NR results for an equal-mass nonspinning
binary. A matching frequency 
is chosen, and at that frequency a phase shift is applied so that the phase
is continuous through the PN-NR connection. There is an amplitude difference
of $\sim10$\% between the PN and NR waveforms (consistent with a 10\% 
uncertainty in the NR amplitude and later calculations of the 6\% error in the 
quadruople amplitude), but the resulting discontinuity is removed from the 
hybrid-waveform amplitude by an appropriate rescaling of the PN component. 
The matching is performed about 12 cycles before merger. 

In \cite{Ajith:2007qp,Ajith:2007kx,Ajith:2007xh} a different procedure was applied. 
PN waveforms modeled
with the TaylorT1 3.5PN phase and quadrupole amplitude were matched with 
NR waveforms with mass ratios $q = M_1/M_2 \in [1,4]$, either with a spacing 
of $\delta q = 0.1$ using short waveforms that include only $\sim4$ cycles 
before merger~\cite{Ajith:2007qp}, or $\delta q = 1$, which include $>10$ 
cycles before merger~\cite{Ajith:2007kx}. The matching is done not at a 
single frequency, but within a time interval, as follows. Consider the NR 
waveform over some 
time interval, $t \in [t_1,t_2]$. Now take a PN waveform with the same
physical parameters (in this case, the mass ratio is the distinguishing parameter),
and align the time and phase of the PN waveform such that the quantity
\begin{equation}
\delta = \int_{t_1}^{t_2} \vert \hc^{PN}(t) - a \hc^{NR}(t) \vert^2 dt,
\end{equation} is minimized. (The constant $a$ is an amplitude scale factor,
and the minimization is also performed over $a$.) Once the PN and NR 
waveforms have been aligned by this procedure, the hybrid waveform is
constructed by linearly interpolating between $\hc^{PN}$ and $\hc^{NR}$ 
over the time interval $(t_1,t_2)$. 

A related procedure is applied in~\cite{Boyle:2009dg}, where a hybrid equal-mass
nonspinning waveform is used to assess the quality of the 2PN stationary-phase-approximation 
(SPA) templates currently used in LIGO searches. Since current
PN templates are cut off before merger, they are expected to be appropriate only for searches
of binaries with a total mass above $\sim25\,M_\odot$ (see, for 
example \cite{Collaboration:2009tt}). The authors 
reach the interesting conclusion that current templates can be used to detect
binaries with much higher masses if the value of the symmetric
mass ratio $\eta = q/(1 + q)^2$ is extended up to unphysical values, $\eta \leq 1$.

In most of the work I have described, NR and PN results were matched between about 
eight and ten cycles before merger. In Sec.~\ref{sec:pn} we saw how accurate
various PN approximants are in this regime, but what does this accuracy mean 
for GW detection? By definition we have no full GR prediction for the PN part of 
the hybrid waveform; if we did we would not need PN theory. The only way to 
assess the accuracy of the hybrid waveforms is to compare hybrid waveforms
produced by different PN approximants at different PN orders. This was done
in~\cite{Pan:2007nw} by calculating the mismatch between hybrid waveforms 
produced using the Taylor approximants at 3PN and 3.5PN order, and between 
3.5PN Taylor and EOB waveforms. If the difference between 3.5PN Taylor- and
EOB-based hybrids is indicative of their {\it physical} error, then these hybrid waveforms
are accurate enough for detection of binaries down to 10\,$M_\odot$ with the 
Initial LIGO, Advanced LIGO and Virgo detectors. If, however, it is the difference
between 3PN and 3.5PN Taylor-based hybrids that we should pay attention to,
then current hybrid waveforms are only useful down to about 20\,$M_\odot$, or
30\,$M_\odot$ in the case of Virgo, which has the broadest frequency range. 
(See Figs.~5 and 6 in~\cite{Pan:2007nw}.) 

Combining the results in \cite{Pan:2007nw} with those of the Samurai 
analysis (Sec.~\ref{sec:consistency}), we see that the accuracy of hybrid waveforms
is dictated by the accuracy of the PN or EOB ingredients. For this reason the
physical accuracy of different PN approaches deserves closer attention in 
the future. 

The construction of hybrid waveforms solves one problem in producing
GW template banks: we now have waveforms of arbitrary length. 
Now we need to produce waveforms for {\it any} black-hole-binary configuration.
One way to achieve this is to devise a general
analytic ansatz for full waveforms, and then use the known hybrid waveforms
as input to determine the unknown coefficients in the ansatz. Such 
``phenomenological'' waveforms have been produced for unequal-mass
nonspinning binaries with
 $q\in[1,4]$~\cite{Ajith:2007qp,Ajith:2007kx,Ajith:2007xh}. 

The phenomenological ansatz used in~\cite{Ajith:2007qp,Ajith:2007kx,Ajith:2007xh} 
is written in the frequency 
domain, and takes the form \beq
u(f) \equiv {\cal A}_{\rm eff}(f) \, {\mathrm e}^{{\mathrm i}\Psi_{\rm eff}(f)}.
\eeq
The effective amplitude ${\cal A}_{\rm eff}(f)$ is made up of three 
piecewise functions: the 
PN quadrupole amplitude, which is proportional to $f^{-7/6}$ in the 
frequency domain, an empirically determined $f^{-2/3}$ behaviour
during plunge and merger~\cite{Buonanno:2006ui}, and finally a Lorentzian is
used to capture the general features of the ringdown. The effective
phase $\Psi_{\rm eff}(f)$ is written as a power series in the frequency.

There are four free parameters in the amplitude ansatz and six in the 
phase ansatz. The set of physical systems that are modeled by this
procedure (i.e., unequal-mass nonspinning binaries) are parametrized
by only two parameters (the total mass $M$ and the mass ratio $q$), but the
ten parameters in the phenomenological ansatz can be mapped to
the two physical parameters by simple quadratic fits. This is an 
important result. It demonstrates that the waveforms (written in terms
of this particular phenomenological ansatz) vary only slowly as the 
mass ratio is changed, and that the phenomenological waveform 
family generated from 31 short simulations separated by $\delta q = 0.1$
can be produced just as well from as few as three simulations 
(i.e., the number of data points needed to fit a quadratic). This allowed
the accuracy of the waveform family to be substantially improved by 
generating four much longer ($>10$ cycles) waveforms with mass
ratios $q = \{1,2,3,4\}$. 

An entirely different procedure to produce accurate analytic waveforms
is based on the effective-one-body (EOB) approach. The original EOB
method~\cite{Buonanno:1998gg,Buonanno00a,Damour:2000we,Damour:2001tu,Buonanno:2005xu}  
predicted waveforms through merger and ringdown before full numerical
simulations were possible, but, despite the compelling case made for
the various physical ingredients in the EOB method, there was no way
to assess the accuracy of its predictions without full GR results to
compare with. 

In the EOB approach the post-Newtonian description of the orbital
dynamics is mapped to an \emph{effective-one-body} description that
incorporates knowledge of the test-mass limit through a resummation of
the post-Newtonian GW energy and flux. This leads to an 
effetive background metric, in which some terms are further resummed
by Pad\'e methods, motivated by physical
arguments~\cite{Damour_T:98}. The resulting inspiral waveforms are
matched to ringdown modes to produce an approximation to the complete
inspiral-merger-ringdown waveform.

Numerical-relativity results have been used to calibrate and enhance
EOB methods in a number of ways. A higher-order 4PN term can be added to
the EOB Hamiltonian and/or flux, and the coefficient(s) fit to numerical
data, first applied in \cite{Buonanno:2007pf}, and later
\cite{Damour:2007yf,Damour:2007vq,Damour:2008te,Damour:2009kr,Buonanno:2009qa},
or some of the existing coefficients in the EOB prescription can also
be fit to numerical 
data~\cite{Damour:2007yf,Damour:2007vq,Damour:2008te,Damour:2009kr,Buonanno:2009qa}. 
Improved matching to the ringdown waveform has also been explored~\cite{Damour:2007yf,Damour:2007vq,Damour:2008te,Baker:2008mj,Buonanno:2009qa}.
One may suspect that the these efforts consist of successively
introducing more parameters until an accurate fit is found. Instead
it has been shown in these works that fitting some EOB coefficients to
only equal-mass nonspinning data then leads to an accurate prediction of
the results for unequal-mass
binaries~\cite{Damour:2008te,Damour:2009kr,Buonanno:2009qa}, and that
the the EOB prescription can be modified to allow accurate fits to
numerical data with only a small number of free
parameters~\cite{Mroue:2008fu,Damour:2009kr,Buonanno:2009qa}. 

Ultimately, one can view the EOB procedure as another form of
phenomenological ansatz. The ingredients in the EOB method are all
physically motivated, but so too is the form of the ansatz used in 
\cite{Ajith:2007qp,Ajith:2007kx,Ajith:2007xh}. In the end it is quite 
reasonable to expect that there exist many different analytic fits to
the full GR waveforms, all of which agree with each other within the
uncertainties in the numerical data or the low-frequency PN
ingredients. The method eventually used to produce templates for GW
searches may depend on which is simpler to implement or, more likely,
historical accident. A more fruitful approach may be to implement
several families of template banks, which allow precise error checking
and comparisons, and may allow those analyzing detector data to more
quickly determine the validity of a detection.

\section{Future issues} 
\label{sec:issues}

This review has focussed on the configurations of black-hole-binary systems
that have been numerically simulated through at least five orbits before merger
and ringdown, and efforts to quantify the accuracy of those simulations, compare
them with PN results, and to combine NR and PN results to produce waveforms
of arbitrary length and phenomenological waveforms that could eventually 
be used to accurately map the entire parameter space of black-hole-binary
waveforms. 

Before this is done, many more numerical waveforms will be needed. This is
a large computational undertaking, and in this final section I will summarize 
a few of the issues that numerical relativists face in producing these simulations. 

{\bf Accuracy of simulations.} Current numerical simulations appear
to be accurate enough for most detection and parameter-estimation purposes with 
current ground-based detectors~\cite{Hannam:2009hh}, at least for
nonspinning binaries. By this I mean that the portion of the 
full inspiral-merger-ringdown waveform produced numerically for a given 
configuration appears to be sufficiently accurate --- whether {\it
  enough} of the full waveform was produced is a different question,
which I discuss below. However, once the Advanced LIGO detector comes
online (around 2013) it is expected to be 10-15 times more sensitive
than current detectors, and as such the accuracy requirements of
waveforms  (in particular for parameter estimation) will increase. A
further increase in accuracy will be necessary to exploit the full
scientific potential of detections from the planned space-based
detector LISA (2018+)~\cite{Danzmann:2003tv}. However, the accuracy
requirements of numerical waveforms have not yet been established for
applications with each of these detectors, and it is important that
they are: not only do we wish to produce waveforms that are accurate 
enough, but we don't want to expend extra effort to produce results
that are more accurate than required --- particularly at the expense
of covering more of the black-hole-binary parameter space. 

There are two main issues in discussing waveform accuracy. One is the 
accuracy of the numerical results from the code. The other is the physical
accuracy of the results, particularly the waveform extraction. It is
the latter (and {\it not} numerical accuracy) that appears to be the
main current bottleneck in improving accuracy, in particular for
higher modes, which will be important for more exotic areas of the
parameter space (see for  
example~\cite{Berti:2007fi,Campanelli:2008nk}), and for more accurate
parameter estimation (see \cite{Babak:2008bu,Thorpe:2008wh} for examples
of estimating the sky location with LISA). 

{\bf Length of simulations.} A second issue, alluded to above, is whether
current simulations are long enough, i.e., include enough inspiral cycles. 
If we wish to use exclusively numerical waveforms as search templates, 
then we have already seen that the simulations need to be hundreds or
thousands of cycles longer --- but few advocate such a severe approach. If we
are to instead connect NR and PN waveforms, as discussed earlier, then we need
to determine that PN waveforms are sufficiently accurate up to the point 
where they are matched to NR waveforms. This is difficult to establish, because
we have no full GR solution to compare against, and if we did, then we could 
use that to match even more cycles before merger. The most fruitful approach
is probably to compare different PN approximants and approaches up to the 
point where the PN waveform would no longer be used. This is a topic that 
deserves greater attention; most studies of NR waveforms prior to the success of 
NR simulations focussed on the faithfulness of templates that used NR
waveforms all the way up to the point where each PN approximant diverges
(see~\cite{Bose:2008ix} for a recent example).  
This grossly over-estimates the discrepency between PN waveforms that
are cut off five, or ten, or more cycles before the standard cutoff frequency.

{\bf Higher mass ratios.} The highest mass ratio binary for which 
full GR numerical results have been published is 
$q = 10$~\cite{Gonzalez:2008bi}. Although the simulations include only 
about three orbits before merger, they are far more computationally expensive
than equivalent equal-mass simulations. The finest resolution of the 
simulation is defined by the smallest mass, while the overall size of 
the binary is defined by the total mass. This means that the computational
cost will scale {\it at best} with the mass ratio $q$. In practice one must
usually provide greater resolution to resolve the dynamics of an unequal-mass
system, meaning that we use a greater number of points on the computational
domain. As such, a 10-orbit simulation of a $q=2$ system that takes six
weeks (as in the simulations used in \cite{Ajith:2007kx,Ajith:2007xh,Damour:2008te})
will take at least fifteen months if repeated for $q = 20$, and could well take
twice that long. Bear in mind that clean error estimates require a convergence
series of at least three simulations, and that we really want waveforms for mass
ratios up to $q=100$ or $q=1000$, or however far we need to go to make
a conclusive connection to the extreme-mass-ratio regime of perturbation theory,
and it becomes clear that we require either phenomenal computer resources,
or a serious advance in numerical methods. 
Exploration of alternative numerical integration methods~\cite{Lau:2008fb} are 
a first step in this direction, although it is likely that more radical approaches
are required, and that the real solution of this problem will represent a 
breakthrough comparable to the first successful black-hole-binary simulations
in 2005.

High-mass-ratio simulations for $q > 100$ are impractical at present, but
this situation could quickly change with improvements in the formulation of
the problem, numerical 
methods and computer hardware. Order-of-magnitude improvements
in either code speed, accuracy or memory efficiency are common when
new numerical methods are introduced. As an example in the opposite
direction, if current typical
mesh-refinement simulations were performed on a uniform grid, the 
memory requirements would be millions of times larger. Full 
black-hole-binary simulations have been possible for only a few years, and the
possibilities for improving on current methods are only beginning to be
explored.

\section*{Acknowledgments}

This work was supported by SFI grant 07/RFP/PHYF148, and I thank the 
Albert Einstein Institute, Potsdam, for hospitality. I am grateful to Sascha
Husa and Niall \'O~Murchadha for many useful discussions, and for a careful 
reading of the manuscript. 

\section*{References}

\bibliographystyle{iopart-num}

\bibliography{refs}

\end{document}